\documentclass[osajnl,twocolumn,amsmath,amssymb,10pt]{revtex4-1} 

\usepackage{graphicx}
\usepackage[english]{babel}

\begin{document}

\title{Ab initio investigation of lasing thresholds in photonic molecules}
\author{Denis Gagnon}
\author{Joey Dumont}
\author{Jean-Luc D\'eziel}
\author{Louis J. Dub\'e}
\email{Corresponding author: ljd@phy.ulaval.ca}

\affiliation{D\'epartement de physique, de g\'enie physique et d'optique \\Facult\'e des Sciences et de G\'enie, Universit\'e Laval, Qu\'ebec G1V 0A6, Canada}

\begin{abstract}
We investigate lasing thresholds in a representative photonic molecule composed of two coupled active cylinders of slightly different radii. Specifically, we use the recently formulated \emph{steady-state ab initio laser theory} (SALT) to assess the effect of the underlying gain transition on lasing frequencies and thresholds. We find that the order in which modes lase can be modified by choosing suitable combinations of the gain center frequency and linewidth, a result that cannot be obtained using the conventional approach of quasi-bound modes. The impact of the gain transition center on the lasing frequencies, the frequency pulling effect, is also quantified.
\end{abstract}

\pacs{140.3410, 140.3945, 230.4555, 140.3430} 


\maketitle

\section{Introduction}

The study of light-matter interactions in photonic molecules (PMs), formed by coupling several optically active microcavities (atoms), has been the object of much work in recent years \cite{Rakovich2010, Boriskina2010}. Applications of microresonators and photonic molecules include optical communications \cite{Griffel1997}, sensing \cite{Boriskina2006, Peng2007, Vollmer2008, Wang2014}, quantum computing \cite{Imamoglu2005} and metrology \cite{Del'Haye2007}. Photonic atoms \cite{Gmachl1998} and molecules \cite{Nakagawa2005} are also well suited for the fabrication of microlasers owing to their high quality factor, or photon recycling rate. These coupled systems also provide a test bed for a panoply of fundamental phenomena including optical bistability \cite{Ishii2006}, coupled-resonator induced transparency \cite{Boriskina2010}, non-reciprocal light transmission \cite{Peng2014} and exceptional points (EPs) \cite{Ryu2009, Liertzer2012}. EPs are ubiquitous in parameter-dependent eigenvalue problems and can lead to surprising physical effects \cite{Heiss2012}. For instance, coupled edge-emitting lasers may turn off even as the pump power \emph{increases} above threshold. This counter-intuitive behavior is achieved by pumping the cavities non-uniformly near an EP \cite{Liertzer2012}.

The lasing characteristics of microresonators are often obtained from the calculation of the cold-cavity (passive) modes. An alternative approach consists in introducing the threshold material gain in the laser eigenvalue problem \cite{Smotrova2006, Smotrova2011, Smotrova2013}. This approach allows assessment of the effect of the resonator geometry on lasing thresholds and emission directionality. However, to take into account the spectral properties of a given laser transition, for instance its position and linewidth, formulations such as the Maxwell-Bloch or Schr\"odinger-Bloch (for 2D systems) theory must be used \cite{Sunada2005, Harayama2011}. This is especially important in the case of nearly degenerate lasing frequencies, for instance near EPs. As we show in this work, the conventional wisdom of cold-cavity modes may not be sufficient in this case.

We will use the \emph{steady-state ab initio laser theory} (SALT) \cite{Ge2010} to assess the effect of gain transition parameters on lasing frequencies and thresholds of a simple two-dimensional PM composed of two coupled active cylinders in the vicinity of a geometric EP. The term \emph{ab initio} refers to the fact that this theory involves the solution of a set of self-consistent equations that explicitly take the gain medium parameters into account. In other words, SALT allows one to determine the steady-state solutions of the Schr\"odinger-Bloch equations \cite{Ge2010th, Harayama2011}. The near-threshold behavior of the PM can be obtained by computing the threshold lasing modes (TLMs) as described in \cite{Ge2010}. This choice of basis states permits to study the effect of the Lorentzian gain transition on thresholds when there are several modes competing for efficient gain extraction. 

This paper is organized as follows. In Section 2, we describe the theoretical background behind our computations, including the main equations of SALT and the method used to compute the lasing states for an arbitrary number of cylinders. In Section 3, we compute the lasing states of a diatomic photonic molecule as a representative example. Our results show that the thresholds of closely spaced modes exhibit a non-trivial dependence on the parameters of the gain transition, specifically the gain center frequency and its linewidth. We also investigate how the lasing modes are subject to the frequency pulling effect, i.e. the effect of the gain center frequency on the spectrum of the PM laser. We summarize our findings in Section 4 and mention a number of possible improvements and extensions.

\section{Theoretical background}

The investigation of the lasing behavior of 2D cavities usually implies the computation of the eigenstates of the passive cavity. These states are governed by the Helmholtz equation
\begin{equation}\label{eq:helm}
[\nabla^2 + \epsilon (\mathbf{r}) k^2]\varphi(\mathbf{r}) = 0
\end{equation}
where an harmonic time dependence $\exp(-i\omega t)$ is assumed and $\epsilon (\mathbf{r})$ is the passive spatially varying refractive index. Both TM $(\varphi \equiv E_z)$ and TE  $(\varphi \equiv H_z)$ polarized waves can be con\-si\-dered. By applying the usual Sommerfeld radiation condition, i.e. an outgoing wave component only, we obtain a set of leaky, or quasi-bound (QB) states characterized by complex eigenfrequencies $k_{\mathrm{QB}} = k' + ik''$ 
\cite{Note1}. 
Since the radiation condition implies $k'' < 0$, QB states are non-orthogonal and exhibit exponential growth towards infinity \cite{Ge2010th}. Despite this unrealistic behavior, QB states provide a useful measure of the photon lifetime in the cavity by means of the quality factor $Q = |k' / 2 k''|$. The $Q$-factor gives a \emph{qualitative} indication of which cavity modes will lase first. Considering a gain transition with center located at frequency $k_a$, one will usually infer that the modes with eigenfrequencies close to $k_a$ and sufficiently high $Q$-factor will lase first \cite{Siegman1986, Sunada2005}.

\subsection{Steady-state ab initio laser theory}

Although the QB states are widely used, they can not describe the lasing behavior of microcavities or PMs in a completely accurate way since the gain medium parameters have no influence on the QB states described by \eqref{eq:helm}. Dynamical theories, for instance the Schr\"odinger-Bloch model, have been used in the past decade to improved the treatment of lasing modes  \cite{Sunada2005, Harayama2011}. In this respect, the recently developed SALT offers an alternative formulation as a stationary version of the Schr\"odinger-Bloch model. Since we are precisely interested in the steady-state lasing behavior of PMs, we shall use SALT in the remainder of this work.

The most important feature of SALT is the introduction of a new kind of eigenstate called a constant-flux (CF) state \cite{Harayama2011, Ge2010, Ge2010th}. The CF states satisfy the following equation
\begin{subequations}\label{eq:cf}
\begin{align}\label{eq:cfa}
[\nabla^2 + \epsilon(\mathbf{r}) K^2(k)]\varphi (\mathbf{r}) &= 0, \qquad \mathbf{r} \in C \\
[\nabla^2 + \epsilon(\mathbf{r}) k^2]\varphi (\mathbf{r}) &= 0, \qquad \mathbf{r} \notin C \label{eq:cfb}
\end{align}
\end{subequations}
where $C$ is the \emph{cavity region}, defined as the union of all optically active regions. CF states are characterized by complex eigenfrequencies $K$ inside $C$, but real frequencies $k$ outside this region. Thus, they exhibit no exponential growth and are physically meaningful, contrary to the QB states \cite{Harayama2011}. 

The laser modes can be expanded in the basis of CF states to take into account effects such as nonlinear coupling between modes (spatial hole burning) and non-uniform pump profiles \cite{Ge2010, Liertzer2012}. For the purpose of this work, we restrict ourselves to the case of near-threshold behavior and spatially uniform pumping inside all cylinders composing of the PM. Under those conditions, the threshold lasing modes (TLMs) of the PM satisfy the following equation
\begin{equation}\label{eq:tlm}
\bigg\lbrace \nabla^2 + \left[ \epsilon(\mathbf{r}) + \dfrac{\gamma_a D_0 F(\mathbf{r})}{k - k_a + i \gamma_a} \right] k^2 \bigg\rbrace \varphi (\mathbf{r}) = 0
\end{equation}
where $k_a$ is the gain center frequency, $\gamma_a$ is the gain width and $D_0$ is associated to the ``pump strength''. Since we suppose uniform pumping of all cylinders forming the PM, $F(\mathbf{r}) = 1$ for $\mathbf{r} \in C$ and $F(\mathbf{r})=0$ for $\mathbf{r} \notin C$ (the cavity region and the cylinders coincide). The TLMs must satisfy an additional reality condition on $D_0$ \cite{Ge2010}. For a given choice of exterior real frequency $k$, the value of $D_0$ is in general complex, but when it crosses the real axis at $k=k_\mu$, we obtain a pair of real numbers $(k_\mu, D_0^\mu)$ defining the TLM lasing frequency and lasing threshold, respectively. The first lasing mode is therefore the TLM with the smallest threshold $D_0^\mu$ \cite{Ge2010}. 

In the case of uniform pumping and a uniform dielectric constant $\epsilon(\mathbf{r}) = \epsilon_c$ inside all cylinders, each TLM can be associated with a single CF state. By comparing \eqref{eq:cfa} and \eqref{eq:tlm}, one obtains the relation
\begin{equation}\label{eq:relation}
\dfrac{\gamma_a D_0}{k - k_a + i \gamma_a} = \epsilon_c \left( \frac{K^2}{k^2} - 1 \right).
\end{equation}
Simply stated, for a given combination of gain parameters ($k_a,\gamma_a$), each TLM of eigenvalue $k_\mu$ corresponds to a CF state of eigenvalue $K_\mu$ for which $D_0$ is purely real, according to \eqref{eq:relation}. 

\subsection{2D Generalized Lorenz-Mie theory}

While SALT allows for arbitrary pump profiles, we restrict our discussion to uniformly pumped dielectric cylinders. This pump uniformity permits the use of the 2D Generalized Lorenz-Mie theory (2D-GLMT) -- also called multipole method -- to compute the eigenstates of the PM \cite{Schwefel2009, Andreasen2011, Note2}. This kind of theory can be used provided the cylinders composing the PM possess enough symmetry to use the method of separation of variables \cite{Gouesbet2013}. It can also be used for the computation of the scattering of arbitrary beams by a complex arrangement of dielectric cylinders \cite{Elsherbeni1992, Gagnon2012a, Gagnon2013}. Multipole-based methods can also be used for modal analysis of quantum dots \cite{Raeis-Zadeh2013}. For the sake of completeness, we review briefly the main equations of the application of 2D-GLMT for the computation of lasing states.

Consider an array of $N$ cylindrical scatterers of radii $u_n$ and relative permittivity $\epsilon_n$. Let also $\mathbf{r}_n = (\rho_n,\theta_n)$ be the cylindrical coordinate system local to the $n^{th}$ scatterer. For modeling purposes, we suppose that every cylinder is infinite along the axial $z$ direction. The central hypothesis of 2D-GLMT is that the total field outside the scatterers can be expanded in a basis of cylindrical functions centered on each individual scatterer, that is
\begin{equation}\label{eq:scat}
\varphi(\mathbf{r}) = \sum_{n=1}^N \sum_{l'=-\infty}^{\infty} b_{nl'} H^{(+)}_{l'} (k_0 \rho_{n}) e^{il'\theta_n}
\end{equation}
where $H^{(+)}_l$ is a Hankel function of the first kind. Inside the $n^{th}$ scatterer, the field can be written as 
\begin{equation}\label{eq:infield}
\varphi(\mathbf{r}) = \sum_{l=-\infty}^{\infty} c_{nl} J_l(k_n \rho_n)e^{il\theta_n}
\end{equation}
where $J_l$ is a Bessel function of the first kind. 

In order to apply electromagnetic boundary conditions at the interface of the $n^{th}$ scatterer, one must find an expression for $\varphi(\mathbf{r})$ outside the scatterers containing only cylindrical harmonics centered on the $n^{th}$ scatterer, that is
\begin{equation}\label{eq:ab}
\varphi(\mathbf{r}) = \sum_{l=-\infty}^{\infty} \left[ a_{nl} J_{l}(k_0 \rho_n) + b_{nl} H^{(+)}_{l}(k_0 \rho_n) \right] e^{il\theta_n}.
\end{equation}
This can be achieved via the application of Graf's addition theorem for cylindrical functions, allowing a translation from the frame of reference of scatterer $n'$ to the frame of reference of scatterer $n$ \cite{Abramowitz1970}. The theorem states that
\begin{widetext}
\begin{equation}\label{eq:graf}
 H^{(+)}_{l'} (k_0 \rho_{n'}) e^{il' \theta_{n'}} = \sum_{l=-\infty}^{\infty}e^{i(l'-l) \phi_{nn'}} H^{(+)}_{l-l'}(k_0R_{nn'}) J_l (k_0 \rho_n) e^{il\theta_n}
\end{equation}
where $R_{nn'}$ is the center-to-center distance between scatterers $n$ and $n'$ and $\phi_{nn'}$ is the angular position of scatterer $n'$ in the frame of reference of scatterer $n$. Substituting \eqref{eq:graf} in \eqref{eq:scat} yields
\begin{equation}\label{eq:graf2}
\varphi(\mathbf{r}) = \sum_{l=-\infty}^{\infty} b_{nl} H^{(+)}_{l} (k_0 \rho_{n}) e^{il\theta_n}
+ \sum_{l=-\infty}^{\infty} \sum_{n' \neq n} \sum_{l'=-\infty}^{\infty} b_{n'l'} e^{i(l'-l) \phi_{nn'}} H^{(+)}_{l-l'}(k_0 R_{nn'}) J_l(k_0 \rho_n)e^{il\theta_n}.
\end{equation}
\end{widetext}
The comparison of \eqref{eq:ab} with \eqref{eq:graf2} then yields the following relation between the $\lbrace a_{nl} \rbrace$ and $\lbrace b_{nl} \rbrace$ coefficients
\begin{equation}\label{eq:anl}
a_{nl} = \sum_{n' \neq n} \sum_{l'=-\infty}^{\infty}e^{i(l'-l) \phi_{nn'}} H^{(+)}_{l-l'}(k_0R_{nn'}) b_{n'l'}.
\end{equation}

A further relation between the $\lbrace a_{nl} \rbrace$ and $\lbrace b_{nl} \rbrace$ coefficients is obtained by applying electromagnetic boundary conditions to \eqref{eq:infield} and \eqref{eq:ab} at $\rho_n = u_n$. This finally leads to the homogeneous equation for the coefficient vector $\mathbf{b}$, $\mathbf{T}(k_n,k_0) \mathbf{b} = 0$, whose non-trivial solutions are given by the condition
\begin{equation}\label{eq:characteristic}
 \det[ \mathbf{T}(k_n,k_0)] = 0
\end{equation}
where $k_n$ is the frequency inside the $n^{th}$ cylinder and $k_0$ is the exterior frequency (both can be complex). One immediately recognizes the transfer matrix $\mathbf{T}$ as the inverse of the usual scattering matrix. $\mathbf{T}$ has a well defined structure; it is composed of blocks containing coupling coefficients between cylindrical harmonics centered on each circular scatterer. Its elements are given by 
\begin{equation}\label{eq:tnn}
\begin{aligned}
& \mathbf{T}_{ll'}^{nn'}  (k_n,k_0)= \delta_{nn'}\delta_{ll'} \\ 
& -(1 - \delta_{nn'}) e^{i(l'-l) \phi_{nn'}} H^{(+)}_{l-l'}(k_0 R_{nn'})s_{nl} (k_n,k_0).
\end{aligned}
\end{equation}
The $s_{nl}$ factor results from the application of electromagnetic boundary conditions and is given by
\begin{equation}\label{eq:snl}
s_{nl}(k_n, k_0) = -\dfrac{J_l'(k_0 u_n) - \Gamma_{nl} J_l (k_0 u_n)}{H^{(+)\prime}_l(k_0 u_n) - \Gamma_{nl} H^{(+)}_l (k_0 u_n)}
\end{equation}
where
\begin{equation}
\Gamma_{nl} = \xi_{n0} \dfrac{k_n J_l'(k_n u_n)}{k_0 J_l(k_n u_n)}
\end{equation}
and $\xi_{ij} = 1 \left( \epsilon_j / \epsilon_i \right)$ for TM (TE) polarization. 
Prime symbols indicate differentiation with respect to the whole argument.
In a typical implementation, $\mathbf{T}$ is composed of $N \times N$ blocks of dimension $2l_{max} + 1$, where $l_{max}$ is chosen sufficiently large to ensure convergence of the cylindrical function expansions. Its value is usually fixed by $l_{max} \geq 3 k \max_n \lbrace u_n \rbrace$. In the case of a diatomic photonic molecule ($N=2$), the matrix is of the form 
\begin{equation}
 \mathbf{T} = 
 \begin{bmatrix}
  \mathbf{T}^{11} & \mathbf{T}^{12} \\
  \mathbf{T}^{21} & \mathbf{T}^{22}
 \end{bmatrix}
\end{equation}
with the diagonal blocks equal to identity matrices and dense off-diagonal blocks, consistent with \eqref{eq:tnn}. More details on the Lorenz-Mie method can be found in \cite{Elsherbeni1992, Nojima2005, Andreasen2011, Schwefel2009, Gagnon2012a, Gouesbet2013, Natarov2014, Note3}.

The QB states of a PM can be computed by substituting $k_n \rightarrow  k \sqrt{\epsilon_n}$ and $k_0 \rightarrow  k \sqrt{\epsilon_0}$ in \eqref{eq:characteristic} and looking for solutions in the complex $k$-plane. As for the CF states, the appropriate substitution is $k_n \rightarrow K \sqrt{\epsilon_n}$ if the $n^{th}$ cylinder is part of the cavity region $C$ and $k_n \rightarrow k \sqrt{\epsilon_n}$ otherwise, with real $k$. The solutions are in this case located in the complex $K$-plane. We note that a countably infinite set of CF states can be computed for each different value of the real exterior frequency $k$.

\begin{figure}
\centering
\includegraphics{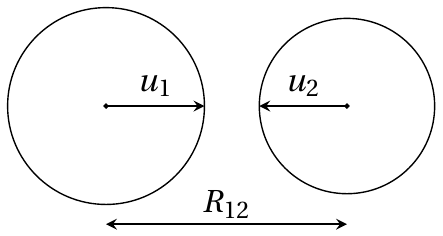}
\caption{Geometry of the diatomic photonic molecule used in this work. The cylinders radii are $u_1$ and $u_2$, with $u_2 = 0.8908~u_1$. The center-to-center distance is $R_{12}=2.448~u_1$ and the relative permittivity of the cylinders is $\epsilon_c = 4$. This geometry is also used in Ref. \cite{Ryu2009}.}\label{fig:pmol}
\end{figure}

\section{Lasing states of a simple photonic molecule}

\begin{figure}
\centering
\includegraphics{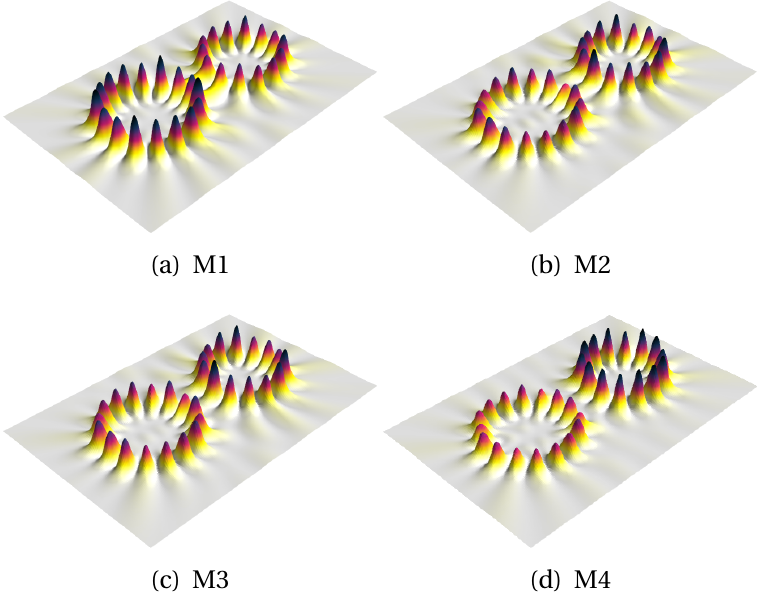}
\caption{(Color online) Profile of four TM-polarized QB states of a diatomic photonic molecule composed of two cylinders of different diameters, as shown in Fig. \ref{fig:pmol}. The $z$-coordinate is proportional to the intensity (arbitrary units).}\label{fig:qbplot}
\end{figure}

\begin{figure*}
\centering
\includegraphics{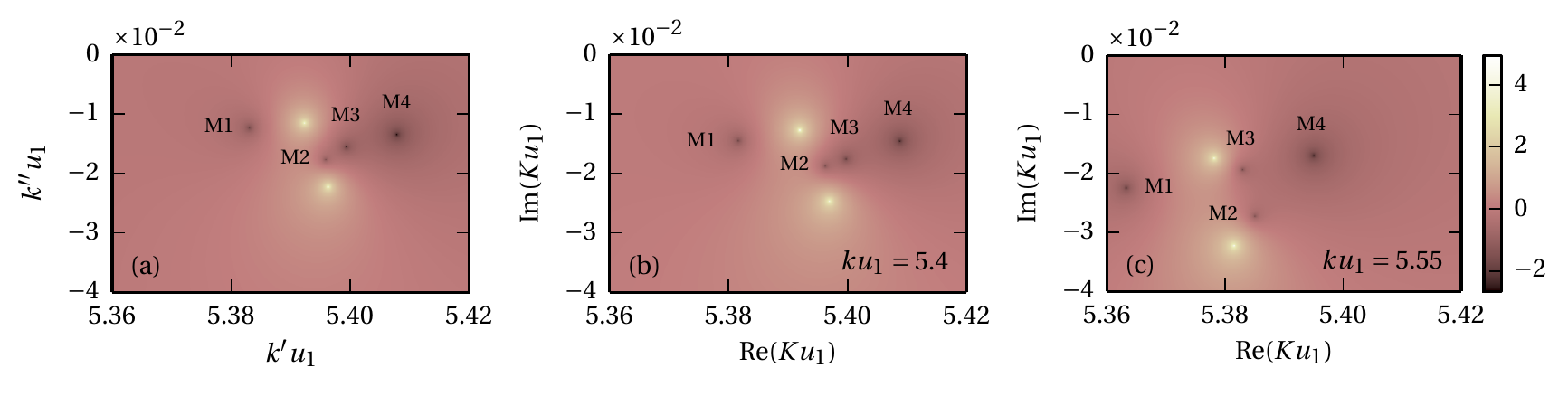}
\caption{(Color online) (a) Map of $\log|\det[\mathbf{T}]|$ in the complex $k$ plane for each of the four QB states of the photonic molecule.
Eigenvalues correspond to the zeros of the function (dark spots) and are located at 
[M1: $k_{\mathrm{QB}} u_1 = 5.3830 - i0.0122$, M2: $k_{\mathrm{QB}} u_1 = 5.3958 -i 0.01756$, M3: $k_{\mathrm{QB}} u_1 = 5.3993 -i0.0154$, M4: $k_{\mathrm{QB}} u_1 = 5.4078 -i 0.0133$]. 
(b-c) Map of $\log|\det[\mathbf{T}]|$ in the complex $K$ plane for two different values of the exterior frequency $k$ (purely real). 
Each QB state can be associated to a unique CF state, allowing the use of the same labels for QB and CF states.}\label{fig:detmap}
\end{figure*}

As a proof of concept, we consider a diatomic PM composed of two coupled cylinders, and restrict the discussion to TM-polarized modes. More specifically, we consider the asymmetric PM as proposed in Ref. \cite{Ryu2009}, shown in Fig. \ref{fig:pmol}. The motivation behind this choice is two-fold. First, it allows us to calibrate the combination of 2D-GLMT and SALT against a previous calculation method, in this case the boundary element method \cite{Ryu2009}. Moreover, the proposed geometry exhibits an avoided crossing of QB states, resulting from of the proximity of a EP of \emph{geometrical nature}. EPs are generically defined as specific parametric combinations for which the eigenvalues of the non-Hermitian operator describing a coupled system \emph{coalesce} \cite{Heiss2012, Liertzer2012}. In the case of the asymmetric PM, the EP can be parametrically encircled by varying the inter-disk distance and the ratio between disk radii \cite{Ryu2009}. Therefore, this choice of geometry is also motivated by the interesting physics of EPs.

The PM is composed of two cylinders of radii $u_1$ and $u_2$, with $u_2 = 0.8908~u_1$. The center-to-center distance is $R_{12}=2.448~u_1$ and the relative permittivity of the cylinders is $\epsilon_c = 4$ (see Fig. \ref{fig:pmol}). The $y$ axis is perpendicular to the line connecting the two cylinders, while the $x$ axis is taken along $R_{12}$.  The near-coalescent states located near $k'u_1 = 5.4$ are shown in Fig. \ref{fig:qbplot}, and the evolution of the associated complex eigenfrequencies is shown in Fig. \ref{fig:detmap}a. The states are split in two symmetry classes with respect to the $x$ axis, odd modes (M1 and M4) and even modes (M2 and M3). They result from the coupling between whispering-gallery modes of slightly different angular momenta of the uncoupled cylinders, creating doublet states \cite{Boriskina2010, Shim2013}.  As a result of the proximity of the geometrical EP, the four QB states located near $k'u_1 = 5.4$ have closely spaced resonance frequencies \cite{Ryu2009}. Moreover, the $Q$-factors are all similar ($Q \sim 200$) and the four states compete for gain. The order in which the modes will lase is not obvious, especially if we consider a gain center with a frequency higher than that of M4. In other words, the conventional approach of considering only $Q$-factors does not allow a quantitative determination of the lowest threshold mode for nearly-degenerate states. Despite this close spacing of modes, we assume that the stationary inversion approximation -- which is central to SALT -- still holds for this geometry. Typical atomic relaxation rates for semi-conductor lasers show that this is indeed the case \cite{Ge2008}.

\subsection{Influence of gain medium parameters}
 
\begin{figure}
\centering
\includegraphics{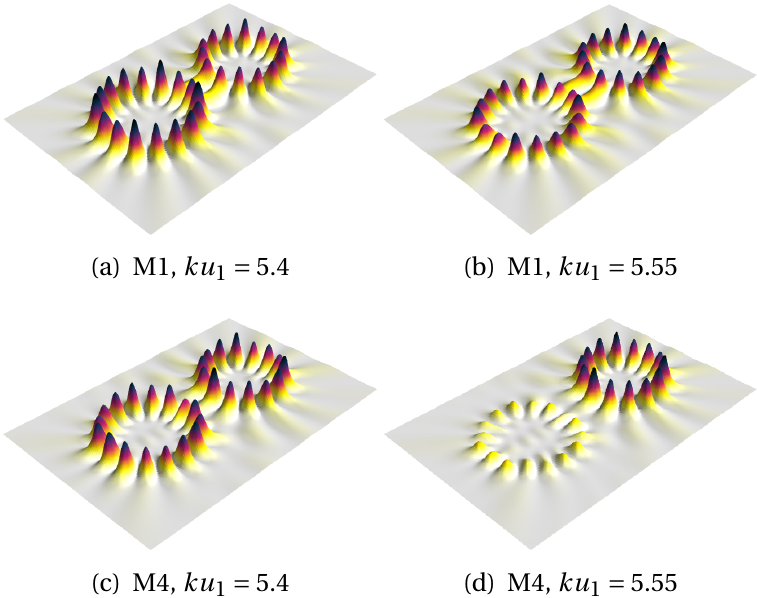}
\caption{(Color online) Profile of two TM-polarized CF states of a diatomic photonic molecule, counterparts to the QB states M1 and M4 shown in Fig. \ref{fig:qbplot}. These profiles are computed for the same values of the exterior frequency used in Figs. \ref{fig:detmap}b and \ref{fig:detmap}c. The $z$-coordinate is proportional to the intensity (arbitrary units).}\label{fig:cfplot}
\end{figure}
 
\begin{figure}
\centering
\includegraphics{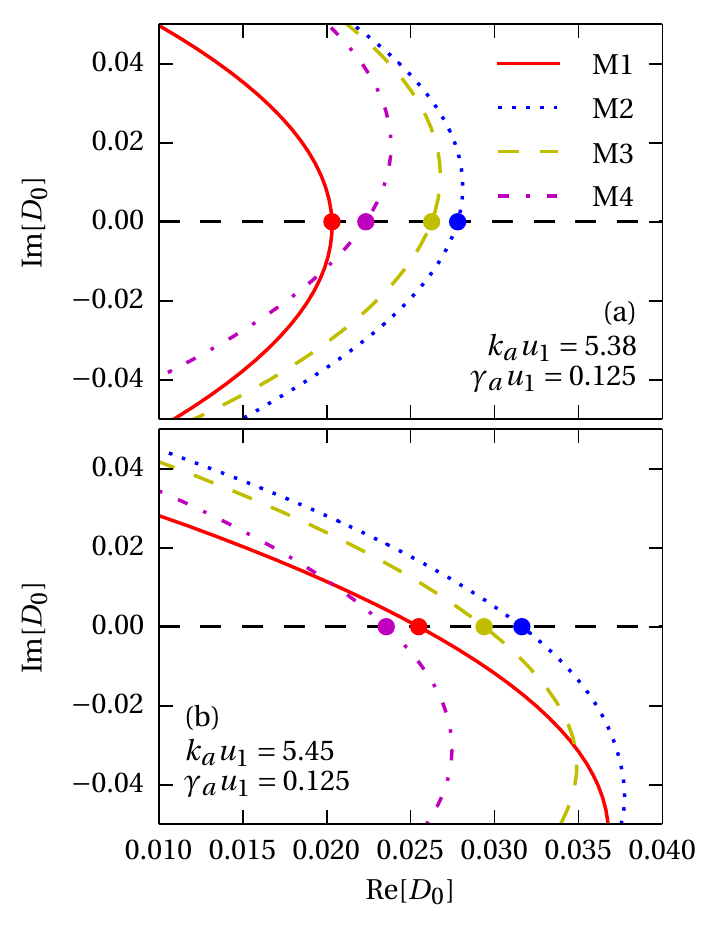}
\caption{(Color online) Evolution of complex $D_0$ values of the four CF states shown in Fig. \ref{fig:detmap} for different values of the gain transition central frequency $k_a$. The threshold for each mode is given by $D_0^{\mu}$ when $\mathrm{Im}[D_0^{\mu}]=0$, indicated by circles on the curves. Note the reversal of M1 and M4 as the first lasing mode when $k_a u_1$ is increased.}\label{fig:dcomp}
\end{figure}
 
\begin{figure}
\centering
\includegraphics{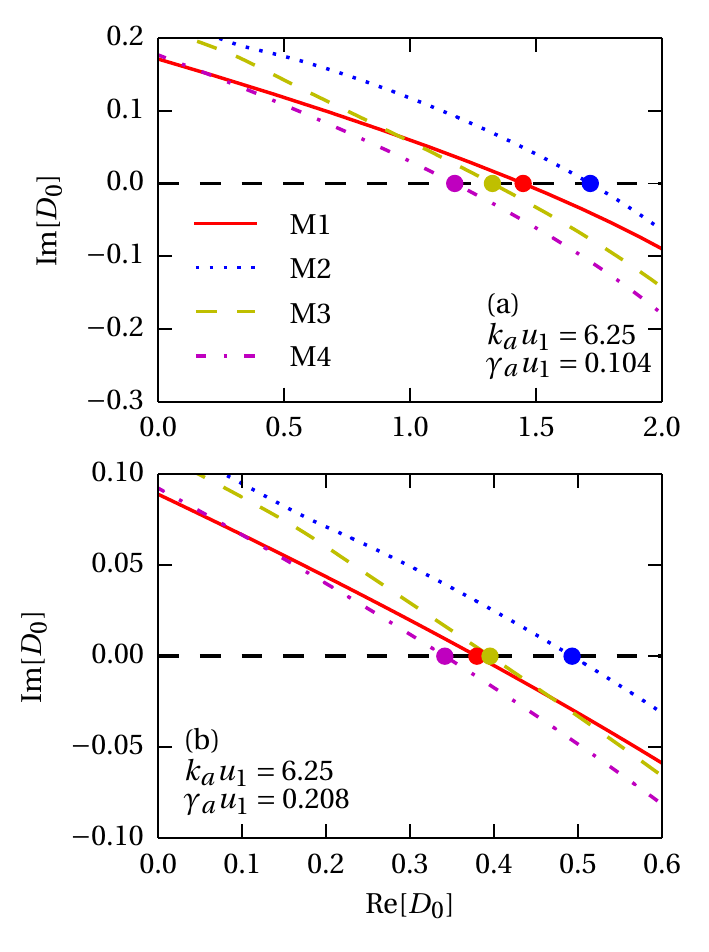}
\caption{(Color online) Evolution of complex $D_0$ values of the four CF states shown in Fig. \ref{fig:detmap} for different values of the gain transition width $\gamma_a$. The threshold for each mode is given by $D_0^{\mu}$ when $\mathrm{Im}[D_0^{\mu}]=0$, indicated by circles on the curves.}\label{fig:dcomp2}
\end{figure}
 
As stated previously, the computation of QB states cannot account for the influence of the gain center frequency and width on the lasing characteristics of a PM. However, if the exterior frequency $k$ is chosen to be close to the frequency $\mathrm{Re}[k_\mathrm{QB}] = k'$ of a QB state, then this state can be associated to a single CF state and their intensity profiles look similar inside the active medium \cite{Ge2010th}. This correspondence can be seen by comparing the symmetries of the amplitude profiles between Figs. \ref{fig:qbplot} and \ref{fig:cfplot}. For other values of the exterior frequency $k$, the intensity profiles may look somewhat different, but the one-to-one correspondence with the QB states still holds (see for instance Figs. \ref{fig:detmap} and \ref{fig:cfplot}). 

Since we restrict ourselves to the case of uniform pumping, each QB state shown in Fig. \ref{fig:qbplot} is also associated to a TLM, making it possible to keep the same labels and compute the associated TLMs to assess the influence of the gain medium parameters. To achieve this goal, one can devise the following procedure for computing the TLM associated to a single QB state
\begin{enumerate}
 \item Compute the complex eigenfrequency $k_{\mathrm{QB}} = k' + i k''$ of the QB state.
 \item Compute the complex eigenfrequency $K(k')$ of the corresponding CF state, using the fact that $K(k')$ and $k_{\mathrm{QB}}$ are usually close \cite{Ge2010th}, as seen in Figs. \ref{fig:detmap}a and \ref{fig:detmap}b for instance.
 \item Compute the values of $K(k)$ in the real neighborhood of $k'$.
 \item Using \eqref{eq:relation}, map the values of $K(k)$ to values of $D_0(k)$. The TLM is characterized by the pair of values $(k_\mu, D_0^\mu)$ for which $D_0$ becomes purely real. An example of this behavior is shown in Fig. \ref{fig:dcomp}.
\end{enumerate}
Once the values of $K(k)$ are computed using \eqref{eq:characteristic}, it is \emph{not} necessary to repeat steps 1--3 when varying the values of $k_a$ and $\gamma_a$ in step 4 as long as the cavity geometry and pump profile are unchanged.

\begin{figure}
\centering
\includegraphics{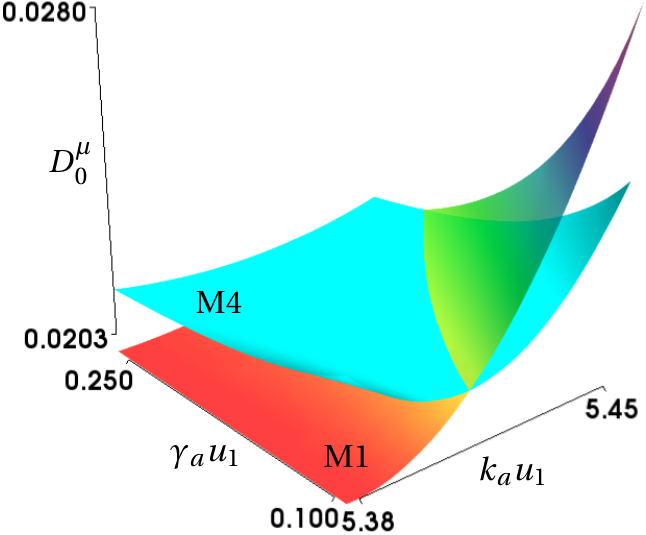}
\caption{(Color online) Evolution of lasing thresholds of modes M1 and M4 as a function of the gain center frequency $k_a$ and gain width $\gamma_a$. 
The thresholds of modes M2 and M3 are higher for this range of parameters (not shown).}\label{fig:thresholds}
\end{figure}

Using this straightforward approach, the dependence of the lasing thresholds and lasing frequencies on the gain medium parameters can be readily investigated for each of the four modes depicted in Fig. \ref{fig:qbplot}. We find that the lowest thresholds modes are always M1 and M4 owing to their higher quality factor. Therefore, we restrict our discussion to these two modes. The dependence of the lasing thresholds $D_0^1$ and $D_0^4$  on $k_a$ and $\gamma_a$ is shown in Figs. \ref{fig:dcomp}--\ref{fig:thresholds}. As expected, M1 is the first lasing mode when the gain center frequency $k_a$ is smaller than the value of $k'$ for that mode. However, as the value of $k_a$ is increased, M1 can still lase first even if $k_a$ is greater than the position of mode M4 (see Fig. \ref{fig:thresholds}). This is especially true for a large gain width $\gamma_a$ as mode M1 is able to extract energy more efficiently from the gain transition in that case, while for a narrow gain transition M4 is favored. 

Interestingly, at the intersection of the two surfaces shown in Fig. \ref{fig:thresholds}, both modes have exactly the same threshold and lase concurrently. Although this lasing behavior is qualitatively consistent with the fact that M1 corresponds to the highest $Q$ cold-cavity mode, the use of SALT is needed to obtain quantitative predictions of its dependence on $k_a$ and $\gamma_a$.

It is also instructive to examine the evolution of lasing thresholds when the gain transition center frequency is far from the QB eigenfrequencies. The dependence of $D_0^{\mu}$ on the value of $\gamma_a$ for a gain transition for large values of $|k - k_a|$ is shown in Fig. \ref{fig:dcomp2}. Our numerical results show that the thresholds of modes approximately quadruple when the value of $\gamma_a$ is reduced by half. Accordingly, one can derive the following expression for $\mathrm{Re}[D_0]$ from \eqref{eq:relation}, under the conditions $\mathrm{Im}[D_0] = 0$ and $|k - k_a| \gg \gamma_a$
\begin{equation}
D_0 \simeq - 2  \epsilon_c \mathrm{Re}[K] \mathrm{Im}[K] \frac{ (k-k_a)^2 }{\gamma_a ^2 k^2}.
\end{equation}
This behavior ($D_0 \sim \gamma_a^{-2}$) is consistent with the observation that modes extract energy more efficiently from a broad lasing transition.

\begin{figure}
\centering
\includegraphics{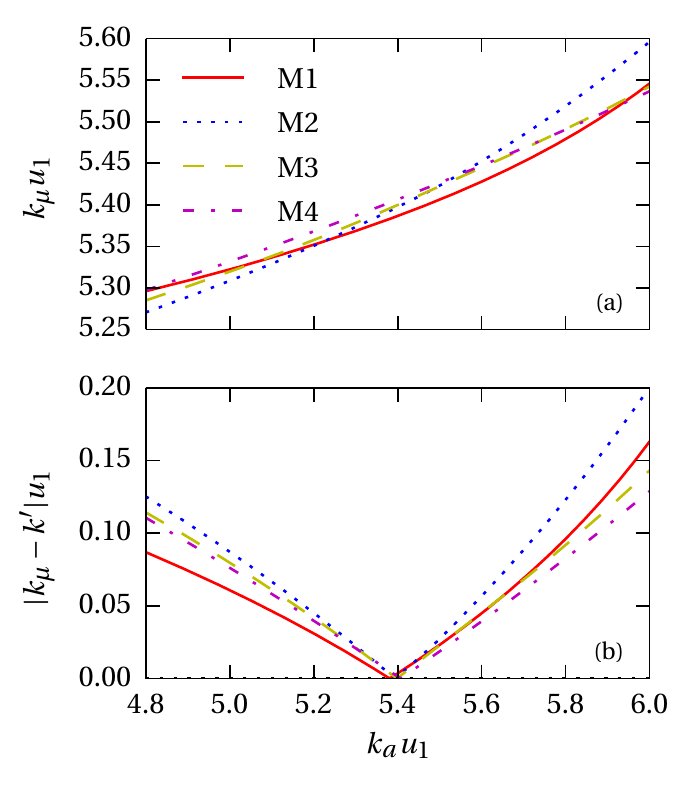}
\caption{(Color online) Frequency pulling effect in a simple photonic molecule for $\gamma_a u_1 = 5.4 \times 10^{-2} $. (a) Evolution of the lasing frequency of each TLM as a function of the gain center frequency $k_a$. (b) Magnitude of the line-pulling effect.}\label{fig:frequencies}
\end{figure}

Next, we assess the influence of the gain center frequency $k_a$ on the lasing frequencies of individual modes. The exact lasing frequency of a mode is always shifted by a small amount from the cold-cavity resonance frequency towards the gain center frequency, an effect known as frequency pulling \cite{Ge2010, Siegman1986}. Since the studied PM geometry exhibits closely spaced lasing modes, this small shift may be of the order of magnitude of the mode spacing. As shown in Fig. \ref{fig:frequencies}a, the order of the lasing frequencies $k_\mu$ can be altered by changing the gain center frequency. For instance, for $k_au_1 \simeq 5.4$, the lasing frequencies are in the order of the QB eigenfrequencies $(1,2,3,4)$, whereas for $k_a u_1 \simeq 6.0$, the order is $(4,3,1,2)$. This can be explained by the fact that lower $Q$ modes are pulled more strongly. As seen in Fig. \ref{fig:frequencies}b, mode M2 is generally the most strongly pulled mode. However, for large values of $k_a$, mode M1 is also subject to strong pulling since it is the mode located further away from the gain center frequency. This result shows that in the case where there are closely spaced cold cavity modes of similar $Q$-factor, for instance in the vicinity of an EP, the spectral characteristics of the PM laser may be strongly affected by the gain transition parameters.

\section{Summary and outlook}

In summary, we have used the threshold lasing modes of SALT to obtain accurate quantitative predictions of the lasing threshold and frequencies of a simple diatomic PM composed of two coupled cylinders. These predictions were obtained from the computation of threshold lasing modes using 2D-GLMT. This combination of SALT and 2D-GLMT is general and not limited to diatomic photonic molecules. For instance, it can readily be applied to the computation of modes of random lasers for an arbitrary number of active scattering centers \cite{Andreasen2011}. 

We found that the lasing thresholds of closely spaced modes of the diatomic PM are strongly influenced by the gain center frequency $k_a$ and its linewidth $\gamma_a$. More specifically the order in which modes lase can be changed by a suitable combination of those gain medium parameters. We also highlighted the frequency pulling effect, and found that lower $Q$ modes are usually subject to stronger pulling. These results show the importance of using \emph{ab initio} theories to take the gain medium characteristics into account in microcavities research. Future work includes an extension to non-uniformly pumped PMs, which precludes however the use of the generalized Lorenz-Mie approach. Although this kind of computations may be achieved via a finite-element method \cite{Liertzer2012}, we have recently developed a versatile scattering approach \cite{Painchaud-April2013} that also allows for inhomogeneous pumping and arbitrary 2D geometries. The generalization to 3D geometries is more challenging, although approaches based on solving the underlying differential equations directly have recently been proposed \cite{Esterhazy2013}. By combining these various numerical schemes with SALT, the path is laid out to investigate, engineer and harness the lasing properties of PMs for ultra-low threshold and directional single-mode emission.

The authors acknowledge financial support from the Natural Sciences and Engineering Research Council of Canada (NSERC). D.G. is supported by a NSERC Postgraduate Scholarship. J.D. and J.L.D. are grateful for a research fellowship from the Canada Excellence Research Chair in Photonic Innovations of Y. Messaddeq. We also acknowledge the free software projects Armadillo \cite{Armadillo} and Mayavi \cite{Mayavi}. Colorblind compliant colormaps are taken from \cite{Geissbuehler2013}. Finally, we thank an anonymous referee for pointing our attention to references \cite{Smotrova2006, Natarov2014}.


\begin{thebibliography}{10}
\newcommand{\enquote}[1]{``#1''}

\bibitem{Rakovich2010}
Y.~P. Rakovich and J.~F. Donegan, \enquote{Photonic atoms and molecules,} Laser
  \& Photon. Rev. \textbf{4}, 179--191 (2010).

\bibitem{Boriskina2010}
S.~V. Boriskina, \emph{Photonic Molecules and Spectral Engineering} (Springer,
  2010), vol. 156 of \emph{Springer Series in Optical Sciences}, chap.~16, pp.
  393--421.

\bibitem{Griffel1997}
G.~Griffel and S.~Arnold, \enquote{Synthesis of variable optical filters using
  meso-optical ring resonator arrays,} in \enquote{LEOS Conf. Proc.},  (1997),
  pp. 165+.

\bibitem{Boriskina2006}
S.~V. Boriskina, \enquote{Spectrally engineered photonic molecules as optical
  sensors with enhanced sensitivity: a proposal and numerical analysis,} J.
  Opt. Soc. Am. B \textbf{23}, 1565--1573 (2006).

\bibitem{Peng2007}
C.~Peng, Z.~Li, and A.~Xu, \enquote{Optical gyroscope based on a coupled
  resonator with the all-optical analogous property of electromagnetically
  induced transparency,} Opt. Express \textbf{15}, 3864--3875 (2007).

\bibitem{Vollmer2008}
F.~Vollmer and S.~Arnold, \enquote{Whispering-gallery-mode biosensing:
  {L}abel-free detection down to single molecules,} Nat. Methods \textbf{5},
  591--596 (2008).

\bibitem{Wang2014}
C.~Wang and C.~P. Search, \enquote{Nonlinearly enhanced refractive index
  sensing in coupled optical microresonators,} Opt. Lett. \textbf{39}, 26--29
  (2014).

\bibitem{Imamoglu2005}
A.~Imamoglu, \emph{{Quantum Computation Using Quantum Dot Spins and
  Microcavities}} (Wiley, 2005), chap.~14, pp. 217--227.

\bibitem{Del'Haye2007}
P.~Del'Haye, A.~Schliesser, O.~Arcizet, T.~Wilken, R.~Holzwarth, and T.~J.
  Kippenberg, \enquote{Optical frequency comb generation from a monolithic
  microresonator,} Nature \textbf{450}, 1214--1217 (2007).

\bibitem{Gmachl1998}
C.~Gmachl, F.~Capasso, E.~E. Narimanov, J.~U. N\"{o}ckel, A.~D. Stone,
  J.~Faist, D.~L. Sivco, and A.~Y. Cho, \enquote{High-power directional
  emission from microlasers with chaotic resonators,} Science \textbf{280},
  1556--1564 (1998).

\bibitem{Nakagawa2005}
A.~Nakagawa, S.~Ishii, and T.~Baba, \enquote{Photonic molecule laser composed
  of {GaInAsP} microdisks,} Appl. Phys. Lett. \textbf{86}, 041112+ (2005).

\bibitem{Ishii2006}
S.~Ishii, A.~Nakagawa, and T.~Baba, \enquote{Modal characteristics and
  bistability in twin microdisk photonic molecule lasers,} IEEE J. Sel. Top.
  Quant. \textbf{12}, 71--77 (2006).

\bibitem{Peng2014}
B.~Peng, S.~K. Ozdemir, F.~Lei, F.~Monifi, M.~Gianfreda, G.~L. Long, S.~Fan,
  F.~Nori, C.~M. Bender, and L.~Yang, \enquote{Parity-time-symmetric
  whispering-gallery microcavities,} Nat Phys \textbf{10}, 394--398 (2014).

\bibitem{Ryu2009}
J.~W. Ryu, S.~Y. Lee, and S.~W. Kim, \enquote{Coupled nonidentical microdisks:
  Avoided crossing of energy levels and unidirectional far-field emission,}
  Phys. Rev. A \textbf{79}, 053858+ (2009).

\bibitem{Liertzer2012}
M.~Liertzer, L.~Ge, A.~Cerjan, A.~D. Stone, H.~E. T\"{u}reci, and S.~Rotter,
  \enquote{{Pump-Induced} exceptional points in lasers,} Phys. Rev. Lett.
  \textbf{108}, 173901+ (2012).

\bibitem{Heiss2012}
W.~D. Heiss, \enquote{The physics of exceptional points,} J. Phys. A: Math.
  Theor. \textbf{45}, 444016+ (2012).

\bibitem{Smotrova2006}
E.~I. Smotrova, A.~I. Nosich, T.~M. Benson, and P.~Sewell, \enquote{Optical
  coupling of whispering-gallery modes of two identical microdisks and its
  effect on photonic molecule lasing,} IEEE J. Sel. Top. Quant. \textbf{12},
  78--85 (2006).

\bibitem{Smotrova2011}
E.~I. Smotrova, V.~O. Byelobrov, T.~M. Benson, J.~Ctyroky, R.~Sauleau, and
  A.~I. Nosich, \enquote{Optical theorem helps understand thresholds of lasing
  in microcavities with active regions,} IEEE J. Quantum Elec. \textbf{47},
  20--30 (2011).

\bibitem{Smotrova2013}
E.~I. Smotrova and A.~I. Nosich, \enquote{Optical coupling of an active
  microdisk to a passive one: effect on the lasing thresholds of the
  whispering-gallery supermodes,} Opt. Lett. \textbf{38}, 2059--2061 (2013).

\bibitem{Sunada2005}
S.~Sunada, T.~Harayama, and K.~S. Ikeda, \enquote{Multimode lasing in
  two-dimensional fully chaotic cavity lasers,} Phys. Rev. E \textbf{71},
  046209 (2005).

\bibitem{Harayama2011}
T.~Harayama and S.~Shinohara, \enquote{Two-dimensional microcavity lasers,}
  Laser Photon. Rev. \textbf{5}, 247--271 (2011).

\bibitem{Ge2010}
L.~Ge, Y.~D. Chong, and A.~D. Stone, \enquote{Steady-state \textit{ab initio}
  laser theory: Generalizations and analytic results,} Phys. Rev. A
  \textbf{82}, 063824+ (2010).

\bibitem{Ge2010th}
L.~Ge, \enquote{{Steady-state Ab Initio Laser Theory and its Applications in
  Random and Complex Media},} Ph.D. thesis, Yale University (2010).

\bibitem{Note1}
Since $\omega = ck$, we will refer generically to both quantities as
  eigenfrequencies.

\bibitem{Siegman1986}
A.~E. Siegman, \emph{Lasers} (University Science Books, 1986).

\bibitem{Schwefel2009}
H.~G. Schwefel and C.~G. Poulton, \enquote{An improved method for calculating
  resonances of multiple dielectric disks arbitrarily positioned in the plane,}
  Opt. Express \textbf{17}, 13178--13186 (2009).

\bibitem{Andreasen2011}
J.~Andreasen, A.~A. Asatryan, L.~C. Botten, M.~A. Byrne, H.~Cao, L.~Ge,
  L.~Labont\'{e}, P.~Sebbah, A.~D. Stone, H.~E. T\"{u}reci, and C.~Vanneste,
  \enquote{Modes of random lasers,} Adv. Opt. Photon. \textbf{3}, 88--127
  (2011).

\bibitem{Note2}
We use the denomination GLMT in accordance with \cite {Gouesbet2013} to mean
  ``theories dealing with the interaction between electromagnetic arbitrary
  shaped beams and a regular particle, allowing one to solve the problem by
  using the method of separation of variables''. However, as a matter of
  historical precision, it could be argued that a theory, dealing specifically
  with the scattering by many cylinders, should be called ``generalized Rayleigh
  theory'' in honour of the first calculation of scattering by one single
  cylinder by Rayleigh \cite{LordRayleigh1881}.

\bibitem{Gouesbet2013}
G.~Gouesbet and J.~A. Lock, \enquote{List of problems for future research in
  generalized {Lorenz--Mie} theories and related topics, review and prospectus
  {[Invited]},} Appl. Opt. \textbf{52}, 897--916 (2013).

\bibitem{Elsherbeni1992}
A.~Z. Elsherbeni and A.~A. Kishk, \enquote{Modeling of cylindrical objects by
  circular dielectric and conducting cylinders,} IEEE Antennas Propag. Mag.
  \textbf{40}, 96--99 (1992).

\bibitem{Gagnon2012a}
D.~Gagnon, J.~Dumont, and L.~J. Dub\'{e}, \enquote{Beam shaping using
  genetically optimized two-dimensional photonic crystals,} J. Opt. Soc. Am. A
  \textbf{29}, 2673--2678 (2012).

\bibitem{Gagnon2013}
D.~Gagnon, J.~Dumont, and L.~J. Dub\'{e}, \enquote{Multiobjective optimization
  in integrated photonics design,} Opt. Lett. \textbf{38}, 2181--2184 (2013).

\bibitem{Raeis-Zadeh2013}
S.~M. Raeis-Zadeh and S.~Safavi-Naeini, \enquote{Multipole-based modal analysis
  of gate-defined quantum dots in graphene,} Eur. Phys. J. B \textbf{86}, 1--7
  (2013).

\bibitem{Abramowitz1970}
M.~Abramowitz and I.~A. Stegun, \emph{Handbook of Mathematical Functions}
  (Dover Publications, 1970).

\bibitem{Nojima2005}
S.~Nojima, \enquote{{Theoretical analysis of feedback mechanisms of
  two-dimensional finite-sized photonic-crystal lasers},} J. Appl. Phys.
  \textbf{98}, 043102+ (2005).

\bibitem{Natarov2014}
D.~M. Natarov, R.~Sauleau, M.~Marciniak, and A.~I. Nosich, \enquote{Effect of
  periodicity in the resonant scattering of light by finite sparse
  configurations of many silver nanowires,} Plasmonics \textbf{9}, 389--407
  (2014).

\bibitem{Note3}
A further technical aspect of the implementation should be noted. Although we
  have not encountered any instabilities in our calculations, it should be
  acknowledged that for high accuracy work, Eqns. (12) and (13) are not well
  suited (exponential decay or growth of $\protect \mathbf {T}_{ll'}^{nn'}$
  with the indices $l$ and $l'$). This difficulty has been recognized before
  \cite {Smotrova2006} and solved generally in \cite {Natarov2014}.

\bibitem{Shim2013}
J.-B. Shim and J.~Wiersig, \enquote{Semiclassical evaluation of frequency
  splittings in coupled optical microdisks,} Opt. Express \textbf{21},
  24240--24253 (2013).

\bibitem{Ge2008}
L.~Ge, R.~J. Tandy, A.~D. Stone, and H.~E. T\"{u}reci, \enquote{Quantitative
  verification of ab initio self-consistent laser theory,} Opt. Express
  \textbf{16}, 16895--16902 (2008).

\bibitem{Painchaud-April2013}
G.~Painchaud-April, J.~Dumont, D.~Gagnon, and L.~J. Dub\'e, \enquote{{S and Q
  matrices reloaded: Applications to open, inhomogeneous, and complex
  cavities},} in \enquote{Transparent Optical Networks (ICTON), 2013 15th
  International Conference on,}  (IEEE, 2013), pp. 1--4.

\bibitem{Esterhazy2013}
S.~Esterhazy, D.~Liu, M.~Liertzer, A.~Cerjan, L.~Ge, K.~G. Makris, A.~D. Stone,
  J.~M. Melenk, S.~G. Johnson, and S.~Rotter, \enquote{A scalable numerical
  approach for the {Steady-State} {Ab-Initio} laser theory,} arXiv.org (2013).

\bibitem{Armadillo}
C.~Sanderson, \enquote{{Armadillo: An Open Source C++ Linear Algebra Library
  for Fast Prototyping and Computationally Intensive Experiments.}} Tech. rep.,
  NICTA (2010).

\bibitem{Mayavi}
P.~Ramachandran and G.~Varoquaux, \enquote{{Mayavi: 3D Visualization of
  Scientific Data},} Comput. Sci. Eng. \textbf{13}, 40--51 (2011).

\bibitem{Geissbuehler2013}
M.~Geissbuehler and T.~Lasser, \enquote{How to display data by color schemes
  compatible with red-green color perception deficiencies,} Opt. Express
  \textbf{21}, 9862--9874 (2013).
  
\bibitem{LordRayleigh1881}
  {Lord Rayleigh}, \enquote{On the electromagnetic theory of light,}
  Philos. Mag. \textbf{12}, 81-101 (1881).

\end{thebibliography}
\end{document}